%% file: paper.tex
\documentclass[conference]{IEEEtran}
\usepackage[normalem]{ulem}
\usepackage{multirow}
\usepackage{url}
\usepackage{verbatim}
\usepackage{graphicx}

\usepackage{soul}    

\begin{document}

\title{
Anception: Application Virtualization for Android
}

\author{
	\IEEEauthorblockN{Earlence Fernandes, Alexander Crowell, Ajit Aluri, Atul Prakash}
	\IEEEauthorblockA{University of Michigan, Ann Arbor\\ (earlence, crowella, aaluri, aprakash)@umich.edu}
	}

\date{}
\maketitle

\input{abstract}

\input{intro}

\input{threatmodel}

\input{overview}

\input{design}
\input{implementation}

\input{functional}

\input{performance}

\input{limitations}

\input{related}

\input{conclusions}

\input{ack}

\bibliographystyle{plain}
\bibliography{reference}

\end{document}

%% file: abstract.tex
\begin{abstract}
  The problem of malware has become significant on Android devices.
  Library operating systems and application virtualization are both possible solutions for confining
  malware. Unfortunately, such solutions do not exist for Android. Designing
  mechanisms for application virtualization is a significant challenge for several
  reasons: (1) graphics performance is important due to popularity of games and
  (2)  applications with the same UID can share state. This paper
  presents Anception, the first flexible application virtualization framework for Android. It
  is implemented as a modification to the Android kernel and supports
  application virtualization that addresses the above requirements.
  Anception is able to confine many types of malware while supporting
  unmodified Android applications. Our Anception-based system exhibits up to 3.9\% overhead 
  on various 2D/3D benchmarks, and 1.8\% overhead on the SunSpider benchmark.
\end{abstract}

%% file: intro.tex
\section{Introduction}
Smartphones have become an attractive target for malware.  The number
of malicious Android apps is growing rapidly and has reached 700,000+
according to a recent
estimate~\cite{TrendLabs_2Q_2013_Security_Roundup}.  Rootkits are a
particularly serious threat since they can completely subvert the
system.

One promising approach to limit the effectiveness of malware is to use
virtualization to help confine malware. One implementation of
virtualization on Android is the Cells~\cite{cells} system. Cells
allows a single Android device to be used as two or more virtual
devices, called ``virtual phones". The user can switch between the
virtual phones using a preconfigured user interface gesture.  It
provides a high-performance approach to support virtual phones. It
does not address the problem of isolating a malware application and a
sensitive banking application on the same phone. Note that in Cells,
each virtual phone is a full Android environment, with its own
capabilities to download and install applications and, possibly, its
own phone number.

Another possible approach to confine malware is the notion of
application sandboxing by using a library operating system (OS). 
Drawbridge \cite{libraryos} is one such system designed for Windows. As
far as we are aware, such a library for Android does not currently
exist and porting Drawbridge to Android will, most likely, be a significant challenge due
to the vast architectural differences between Windows and Android.
More importantly, there are three fundamental drawbacks to using a Drawbridge-like
solution on Android. First, Drawbridge uses RDP, Windows Remote Desktop Protocol, for
virtualized applications to render graphics on the host system. This approach
is likely to be impractical on Android devices because many popular applications,
especially games, are graphics intensive. It will also impact the display's
responsiveness to user input. 

Second, Drawbridge does not support shared state between
applications. Each application is a sandboxed process and sharing
state (e.g., files) among processes remains an unresolved problem.
Android permits such sharing among applications with the same UID. The
UID of an application, in Android, distinguishes the origin of the
application, e.g., the vendor and the associated package to which the
application belongs.

Finally, Drawbridge does not support multiple processes bound to a
single OS library.  This is a significant limitation. Fork/clone need
to be supported on Android.

This paper describes a novel virtualization approach, called
Anception, that provides flexible application virtualization on Android,
addressing the drawbacks of previous approaches.  Anception provides
the following features: 

\begin{itemize}
\item A notion of \textit{containers} that provide a lightweight virtualized OS to applications.
\item Efficient graphics (close to native speed), with native graphics
acceleration capabilities, to all applications. 
\item Support for binding multiple applications to the same container so
that they can share filesystem state, file handles, etc. Furthermore, each container
can support multiple processes. Fork/clone are supported.
\end{itemize}

Anception permits sharing among applications under an appropriate
policy.  A reasonable policy for the use of Anception is to only
allow applications with the same UID to share the container. This
would be consistent with the Android security model and would prevent a
banking application, provided by a bank, from being in the same
security container as a gaming application, provided by an unverified
developer. At the same time, the policy would permit the bank app to
co-operate with other trusted applications (possibly by the same developer) that share a common persistent
state on the file system. With the above policy, Anception does not change the current
Android experience.

Anception's design also provides higher resiliency to a sensitive app
from a buggy and possibly malicious app when they are assigned to the
same container. A scenario is where a bank provides two apps that
share a package name or a shared uid field. One of the apps is for
online banking and another is a financial analysis package that may
access a common local database between the two apps for calculating
balances, plotting trends, etc. The online banking application needs
access to the user's online banking password and is very carefully
designed to only keep the password in memory for the minimum time and
always send it encrypted to a remote system.  The financial analysis app
on the other hand, is not as carefully designed and provides an
exploit path so that it can be used by an attacker to acquire elevated privileges on the container.
For this scenario it would normally be difficult for the banking app to protect its secrets from
the financial calculator and yet be in the same container to share system state. 
Anception's memory isolation design allows the banking app to better safeguard the secrets in memory from 
the compromised financial analysis app.

The key contributions of this paper include:
\begin{enumerate} 
\item An application virtualization mechanism for the Android kernel that provides
good performance for graphics-intensive workloads and allows applications in 
the same virtual machine to share filesystem state. 
The solution does not require any changes to existing Android applications
\item Memory isolation among 
processes in the same container, as provided by our application virtualization mechanism. This guarantee is upheld even when
the container is compromised.
\item An efficient system call interposition method to transfer system calls
to an application's virtual machine
\end{enumerate}

To provide the above contributions, Anception uses a novel form of
virtualization which we call {\em headless virtualization}. With this
form of virtualization, security {\em containers} for applications are
implemented as headless virtual machines, i.e., they do not have the
user interface inside the virtual machine; instead the user interface
continues to reside on the host, providing close-to-native graphics
performance. In Anception, a virtualized application launches from the
host but relies on its container to service most of the system
calls. This reduces the attack surface available to the application to
compromise the system.

Anception is an x86-based prototype, consisting of approximately 5500
lines of code. It requires minimal changes to the Android kernel. Most of the
code resides in two Linux kernel modules, one for the host
system and one for implementing the containers using guest virtual machines. 
We created a headless Android userspace stack for the guest virtual machines. 

We evaluated an Anception-based system with respect to both security
and performance. We analyzed previously reported vulnerabilities in
Android and found that Anception provides comparable protection to
solutions based on full virtualization. With respect to performance,
we analyzed Anception using on several popular benchmarks. Anception
incurs a 1.2\% overhead on the SunSpider benchmark and less than 3.9\%
overhead on a variety of PassMark 2D and 3D graphics benchmarks. Power
consumption overheads of the Passmark benchmark running within an
Anception container was 2.4\%. Because of the headless design,
Anception containers are memory efficient. They can be launched with
44MB of memory assigned to them versus the 55MB for
Cells~\cite{cells}.

Rest of the paper is structured as follows. In section \ref{threatmodel} we present the 
threat model for Anception based systems. Section \ref{sec:anceptionmodel} presents the 
high level overview of the Anception model and Section \ref{sec:design} presents the design 
and implementation details. In section \ref{sec:evaluation} we present a security evaluation 
of our model and in section \ref{sec:experiments} discuss the performance overheads. Section \ref{sec:limitations}
discusses the limitations of our approach. We present the related work in section \ref{related} and 
conclude in section \ref{sec:conclusion}.

%% file: threatmodel.tex
\section{Anception Threat Model}
\label{threatmodel}

Our current implementation assumes that the base set of applications
that are preinstalled on an Android device, e.g., Photo Gallery and Contacts Manager, can
be trusted.  In principle, they could all be encapsulated in
Anception containers for additional security, but our current
implementation does not do that.

We assume that the host operating system kernel is trusted and cannot
be compromised by a virtualized untrusted app. Except for display (and touch)
management services, we provide virtualized equivalents of various
privileged Android services (e.g. LocationManagerService, vold) that execute in a container. 
Untrusted apps may attempt to exploit them to acquire elevated
privileges. We assume that a malicious app may completely compromise its container.

We assume that apps may attempt to open \texttt{/dev} devices and issue I/O control
operations on them in an effort to exploit bugs in the kernel.

Anception assumes that an application is capable of acting in
malicious ways.  For example, it can simply ask for more permissions than
needed and use them to exfiltrate user data. An application can also
execute native code on the device including system calls allowed by
the operating system security policy (which is usually most system calls).
This often takes the form of local privilege escalation attacks that exploit a
vulnerability in some privileged component or
framework code. 

Finally, our threat model also includes combinations of these
techniques wherein a benign application may download an exploit and execute it
to gain elevated privileges. It then proceeds to install backdoors and steal
user data. However, we cannot protect against all types of privilege escalations.
Exploits that involve direct exploitation of a kernel interface
are outside the scope of our threat model. 

Covert channels are also outside the scope
of this paper.

Hardware devices are ultimately shared among all containers. Low-level
exploits on the device drivers in the host are outside the threat
model.  Blocking access to devices, e.g., sensors, selectively for
containers is possible future work, based on prior work in the
area~\cite{Beresford:2011:MTP:2184489.2184500,Nauman:2010:AEA:1755688.1755732,Xu:2012:APP:2362793.2362820}.

%% file: overview.tex
\section{Overview of Approach}
\label{sec:anceptionmodel}

\begin{figure}[ht!]
\center
\includegraphics[width=3.3in,height=2.8in]{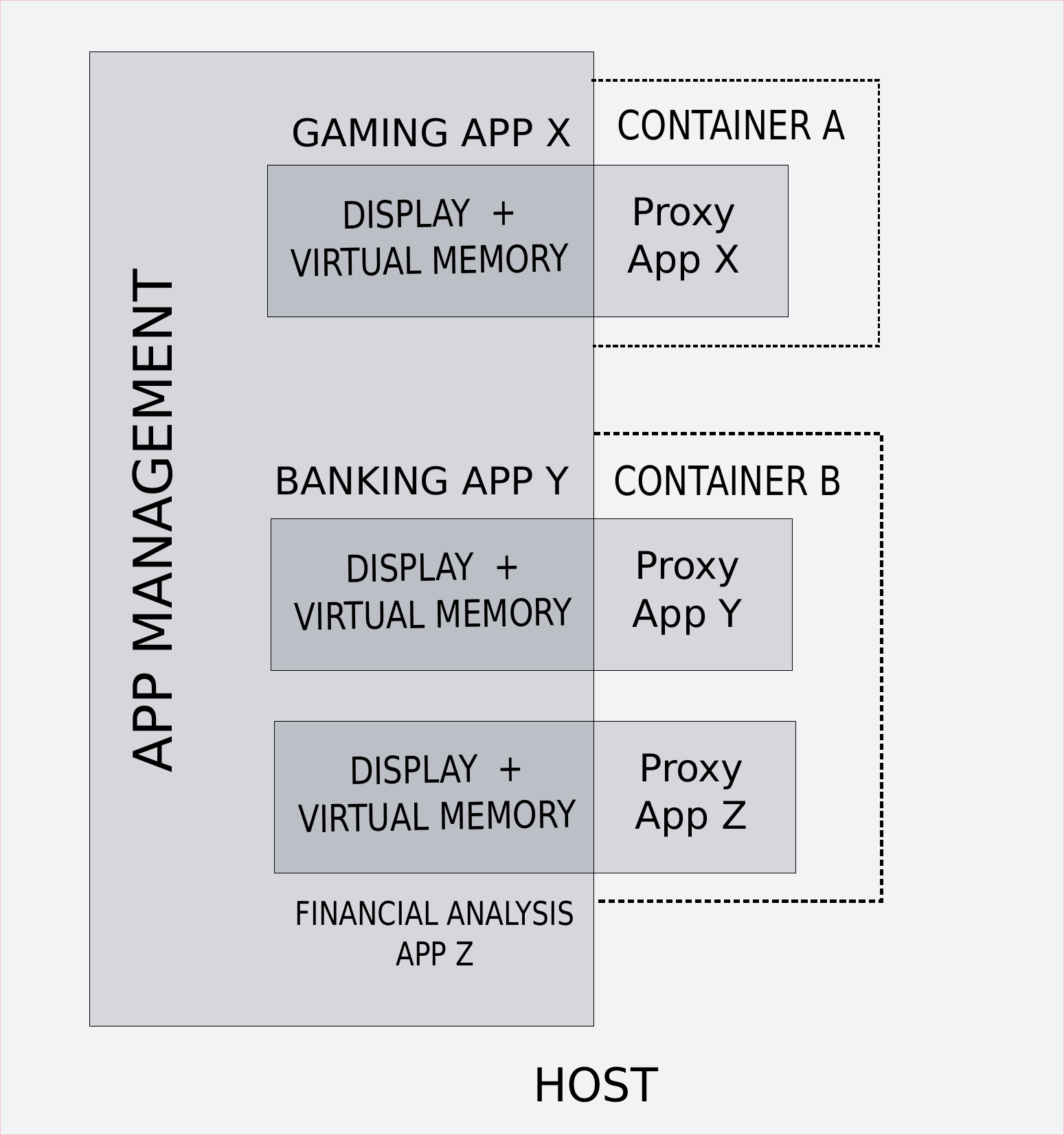}
\caption{Model Overview}
\label{fig:modeloverview}
\end{figure}

The Anception model strengthens isolation between groups of smartphone
applications and between applications and the host OS by providing a
notion of containers (see Figure~\ref{fig:modeloverview}). Containers
in Anception are implemented by using {\em lguest}\cite{lguest}, a lightweight type II
hypervisor.  Containers provide resilience against malware that execute
privilege escalation attacks. Those attacks, in most cases, are restricted to their container. 
Containers also provide resource isolation. Apps in different containers do not share
filesystem state.

Figure~\ref{fig:modeloverview} shows the relationship between apps and
containers.  A container may be associated with one or more apps. A
reasonable system policy is to associate apps sharing a UID with the
same container. In Android, apps with the same UID have the same
rights and can share filesystem state. We assume this policy.
Figure~\ref{fig:modeloverview} shows three user apps. App
$X$ is bound to container $A$ and apps $Y$ and $Z$ are both bound to
container $B$. App $X$ can exchange data within the
device with app $Y$ and $Z$ via IPC.  Apps $Y$ and $Z$ can share system state such as files 
with each other and thus have additional sharing channels since they are in the same container.

A downloaded app is bound to a container.  The container maintains
persistent state of that app. This primarily includes the filesystem
state. It also includes the kernel state associated with the app,
e.g,. state of open files, open network or IPC sockets. 

A pre-installed app that comes with the phone need not be bound to a container. 
As indicated in Section~\ref{threatmodel}, we trust pre-installed apps not
to be malicious. If additional resilience is desired, pre-installed apps could
be put in a container.

IPC messages between apps, even across container boundaries, are
allowed to maintain compatibility with the Android sharing model. In
Android, apps can send requests to each other via {\em intents}, which
are implemented using \texttt{ioctl} calls on \texttt{/dev/binder}. The
system then delivers IPC messages to the recipient(s), subject to
access control rules. Recipients can choose to accept or reject the
requests. This allows apps to cooperate with each other. For example, Snapchat
app can request the Photo Gallery to take a photo. Then, it can request the Facebook
app to upload the photo to a user's Facebook page. Allowing these IPCs is crucial
to the Android model.

Providing fast graphics for virtualized apps is generally considered
difficult.  To allow virtualized apps to get close to native graphics
performance, we chose to virtualize an app partially. The app launches
normally in the host. Its memory pages reside on the host. User-space
instructions execute on the host. Most system calls, including most
\texttt{ioctl}s, are directed to the app's container. They are executed by an
app's proxy process in the container. This provides fast display
updates and interactive performance at a potential expense of slower
filesystem calls.  An alternative design would have been to do the
graphics operations in the container and use a remote windowing
protocol such as X windows, VNC, or RDP. In that case, filesystem
operations would be fast but graphics operations would be
slower. We provide experimental data in the section on Evaluation that
supports our thesis that optimizing user-interface performance is more
important than filesystem performance for most apps.

The above design constraint for speeding up graphics implies that all apps
must launch from the host. An alternative design is to launch the app from the container
but that would have slowed down the performance of \texttt{ioctl}s, which are used
for sending intents and display-related requests. Those \texttt{ioctl}s would have to
go back to the host for sending to apps in other containers or for updating the display.
That would imply a big context switch from the container's VM to the host. We avoid most
of these context switches by launching the apps from the host. As Section~\ref{sec:evaluation} 
substantiates, \texttt{ioctl}s are very frequent on Android as compared to other system calls

Given that apps launch from the host, the virtual memory pages of apps reside on the host.
But most system calls, e.g., file writes,  are carried out on its container's kernel.
This implies that a container's kernel cannot directly inspect the memory pages of an app
that uses it. Conceptually, the container is like a server to the app and manages
its system state. This provides an additional layer of protection to apps. 
We described a scenario in the Introduction where a banking app and a financial analysis 
app need to be in the same container. As long as the banking app does not write
the user's password to the disk or send it unencrypted over the network, the financial
analysis app cannot get to it, even if it is a rootkit. The financial analysis app may
be able to totally subvert the shared container, but the container does not have the ability
to inspect the banking app's virtual memory. The container could attempt to core dump the
app, but that would cause the host operating system to be the one to dump code. 
That will be written to the host OS. The container could also attempt to return bad data in
system calls to try to do Iago attacks~\cite{Checkoway:2013:IAW:2451116.2451145}. We are not
able to prevent those; the banking app must be designed to be resilient to those attacks. Any critical
certificates or keys used by the app must also be part of its code (which are stored
on the host side) so that they cannot be tampered with by the container.

An app should only be able to write to its container. But, it is acceptable
for it to read the read-only part of the file system from the host. 
The following principle is used for directing VFS (virtual file
system) layer operations for disk files: 
\begin{itemize}
\item Access to read-only parts of the file system, e.g., {\tt /system}, application code, and system
libraries, are directed to the host
\item Others are directed to the container. 
\end{itemize}

The above principle is important to protect the banking app
from the malicious financial analysis app when they are in the same container. 
If the code or system libraries were read from the container's filesystem, the malicious app
could modify them and corrupt the banking app.

The device filesystem provides access to various hardware or virtual
devices.  Some examples are the framebuffer, the accelerometer and
Binder. On an Android system, all hardware devices are managed by the
system. Thus no app should be able to directly communicate with the
device files for these devices. So we just forward these
requests to virtual devices in the container. The two exceptions to this
policy are {\tt /dev/binder} and {\tt /dev/ashmem}. Access to these
devices is needed by apps to support IPCs and shared memory.  Access to
these is directed to the host.

In Anception, each container has instances of standard Android system
services. For example, one of the standard system services provides
access to the contacts database. An app in a container should use the
container's contacts database. These services are invoked via
IPCs. This implies that such IPCs directed at system services,
must be redirected to the container.

One exception to the redirection of IPCs to system services is the
Notification Manager service that notifies users of notifications from
apps. Since user-interface is provided by the host, any IPCs to the
Notification Manager must be delivered to the host, rather than
to the container.

In Anception, all apps are always installed on the trusted host and
launched from there, even though apps are bound to their respective
containers. That implies IPCs between two apps that are bound to the
same container must be delivered via the host, i.e., such
IPCs must not be redirected to the container.

%% file: design.tex
\section{Anception Implementation}
\label{sec:design}
Our prototype consists of two Linux kernel modules (one host, one guest), a few
userspace helpers, and a headless Android userspace stack. We use the 2.6.39
x86 Asus EEEPC kernel, Android 2.3 Gingerbread for x86 and an Asus EEEPC ET1602
tablet for the prototype. The implementation is a little over 5500 lines of
code.  Our changes are minimal and only add a few lines of code to the system
call handler, the process descriptor and the fork system call (approximately 30
lines in total).  The rest of the code is within kernel modules.  The lguest
hypervisor is approximately 6000 lines of code. Below, we describe the design
of the system.

\begin{figure}[ht!]
\center
\includegraphics[width=3.5in]{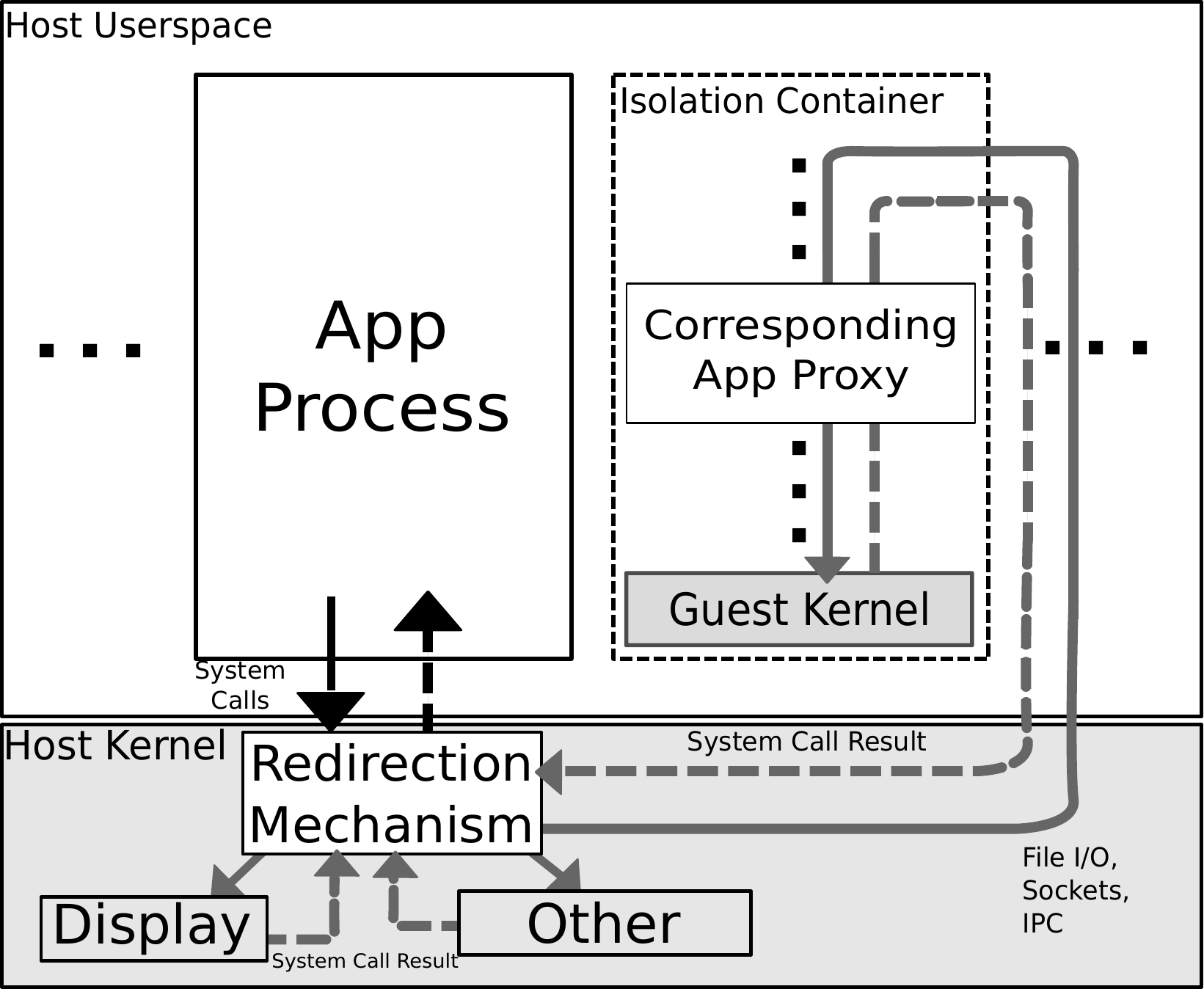}
\caption{High-level Anception Architecture: An app has a proxy in the
container bound to the app. System calls that need to be confined are
redirected to the proxy.}

\label{fig:arch2}
\end{figure}

\begin{figure}[ht!]
\center
\includegraphics[width=3.5in]{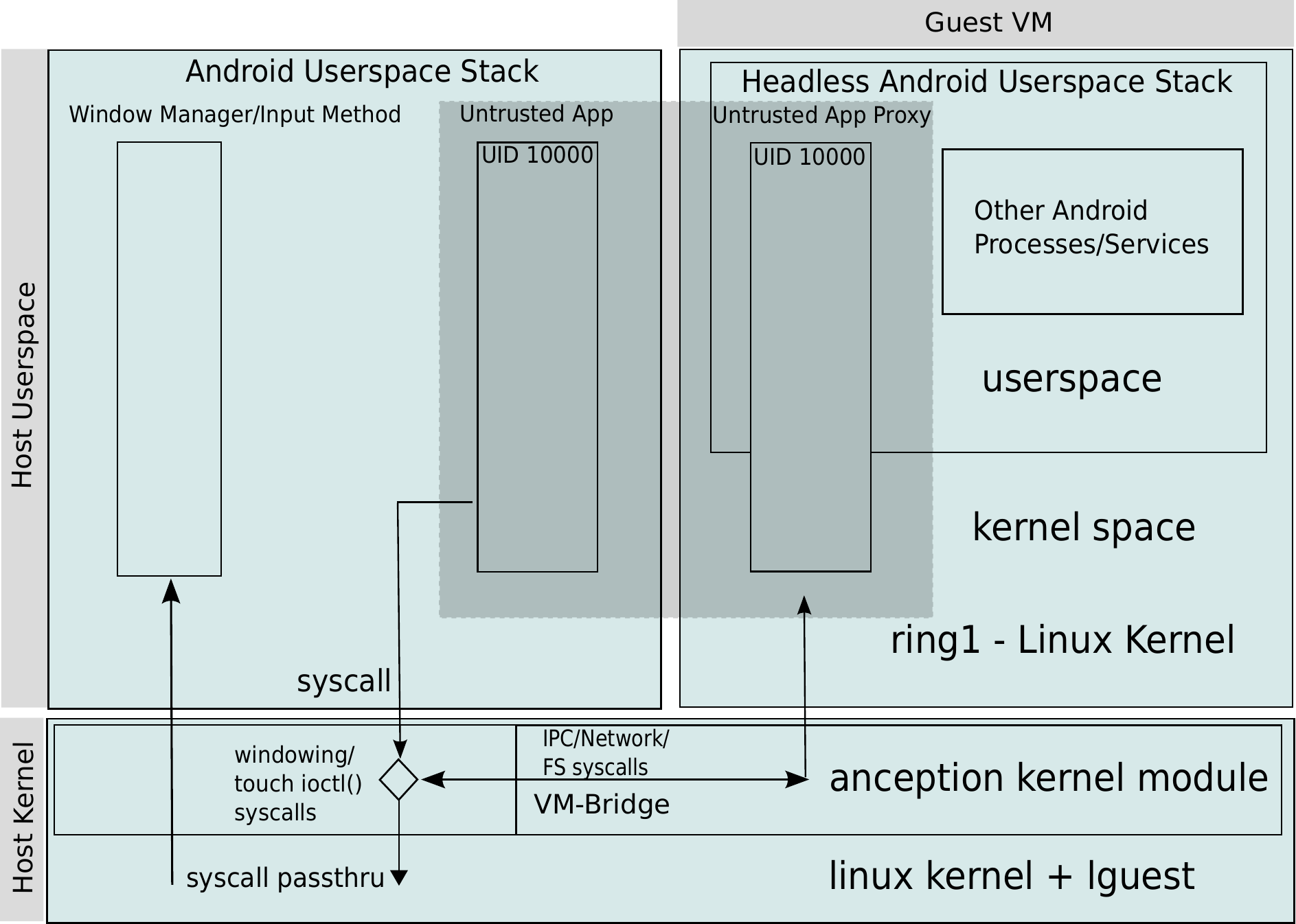}
\caption{Anception Architecture: The Anception kernel module contains the VM independent (left) and VM dependent portions (right)}
\label{fig:arch1}
\end{figure}

Figure~\ref{fig:arch2} shows the high-level architecture for
implementing Anception containers and Figure~\ref{fig:arch1} shows
further details. Each Anception container corresponds to a
minimalistic guest virtual machine to provide isolation
properties. Apps launch in the trusted system but system calls that
could potentially break the isolation properties on Android are
intercepted and redirected to the guest virtual machine. On our
Android-based implementation, this includes all file-system related
calls and most \texttt{ioctl} calls depending upon the level of sharing
we wish to configure.

As stated in the previous section, all the user-interface aspects of
the app remain in the trusted host system for all the
containers. Unlike desktop-based systems, Android already provides
strong isolation at the user interface level between apps and we rely
on that. Besides speeding up graphics, another advantage of this is
that it allows each guest virtual machine to have a small memory
footprint.

The host kernel runs a normal Android stack. Applications are bound to
guest kernels representing the various containers. The Anception
container runs a software stack identical to the host stack in most
respects, except in its ability to render to the display or process
incoming touch input. That is, the guest is an identical, {\em
  headless} version of the host software stack. The identical portions
include the kernel image and framework code that forms an Android
runtime.  

When an application is launched, a proxy process is launched inside
its container VM as well (see Figure~\ref{fig:arch1}). The proxy
process sees equivalent privileges, UIDs and directory structures
within the guest.  The purpose of the proxy is to execute forwarded
system calls and return the results. When the process dies, a request
is sent to kill its proxy as well.

The architecture of Anception poses a design challenge when it comes to
managing memory. An app first launches in the host. At the same time,
a proxy process is created in the container to which the app belongs.
The app's page tables and user-space memory resides on the host.
But, when it makes a system call, somehow the parameters of that call must
be made available to the proxy since the proxy is supposed to execute the
call. As a concrete example, consider a system call to open a file. The
file has parameters such as the name of the file to be opened. If the
system call were executing locally, the kernel can grab the parameters
by reaching into app's user-space memory. In Anception's case, the
call is forwarded for execution on a remote kernel. That guest kernel does not
have the ability to get to app's user-space memory.

To address the above problem, Anception marshals all the
parameters of the call (in a similar fashion to the seccomp sandbox \cite{seccomp})
 and sends them to the proxy. For the above call,
the filename would be marshalled and sent to the proxy. The proxy opens
the file on its own kernel and gets a file handle. The file handle is 
returned back to the host system. 

Note that system resources, such as actual files and their file
handles, exist within the container. The app on the host side gets
a handle to a remote resource. That handle can be used in subsequent 
read/write calls.

Marshalling occurs on write calls as well. Data that is
written to a file must be marshalled and sent to the
proxy. Fortunately, marshalling turned out to be straightforward for
most calls. Also marshalled data for most calls, except for a few such
as large file writes, consist of only small amount of data. In a
naive implementation, additional data copies would be required -- 
from host kernel to the proxy, from the proxy to the container's kernel,
return of the result from the container's kernel to the proxy process, and then marshalling of the result from the proxy process to the host kernel. 

Fork/clone calls need to be handled carefully. It is not sufficient to
fork/clone an app's process since its file handles really reside with
its proxy. Fortunately, the problem is easily solved by also
forking/cloning the proxy. So, in reality, each process within an app
is bound to its corresponding proxy. Note that the container ID
is always inherited whenever these operations take place. A process cannot escape 
its container through forking/cloning.

The \texttt{execve} call is handled in the following way.
First of all, nothing special needs to be done to the proxy on an {\em
  exec} system call. The proxy continues to store the resource
handles. The host process executes the new code. If the binary being
executed is a system binary, it is simply executed on the host since
the host's version is identical to the guest's. If it is a user
generated binary, it is copied out from the guest to a special
execution cache that is not accessible to the untrusted app, and
executed from there. The reason for this is that we don't want the app
to trick the system into copying an executable to a restricted
location. Expressing the policy this way is much cleaner. Note that
any newly executed binary will belong to the same container as that
of the parent.

Dangerous calls like \texttt{insmod, rmmod, shutdown} and others
relating to whole system management are denied to applications because
no user downloaded app should ever invoke these. Android security model
denies them as well.

Android allows a process to create shared memory segments that can be
mapped into memory of multiple processes. Handling shared memory
segments poses interesting design options. The question is whether
these should be created within the host process or in the proxy.  The
problem is that shared memory segments are often mapped into virtual
memory after creation and accessed from code (via the \texttt{mmap}
system all).  Creating these shared memory segments in the proxy would
have required intercepting read/write memory operations to the pages
corresponding to shared memory segments and transferring them to the
proxy, which would not be very efficient.

To address the above problem, mechanisms such as distributed shared
memory~\cite{li98,Keleher:1994:TDS:1267074.1267084} could have been
used, but we chose to go with a simpler solution.  We allow shared
memory segments to be created on the host. This may appear to
introduce a security risk if memory segments could be shared between
processes belonging to different containers. Fortunately, that sharing
can be easily prevented. In Android, unlike Linux, memory segments
cannot be shared by simply knowing the segment name. Instead, Binder
IPC calls must be used to communicate memory segment
handles. Anception restricts such IPC calls to processes within the
same container. Thus, apps in different containers cannot share memory
segments.

%% file: implementation.tex
\subsection{Management of Apps and Containers}
All Android apps are started by a fork from the Zygote process (which is like {\em init} in Linux). 
After the fork, Zygote specializes the child into an application by setting
its UID and other parameters. We added code to notify the Anception
kernel module of this, and we replicate the same on the guest by
spawning a proxy, and specializing it.  A userspace daemon executes on
the host and is responsible for the VM management functions like startup
and shutdown. It communicates with the host module through a
miscellaneous device node. It is also responsible for binding an
application to a container by setting its \texttt{vmid} flag to a
valid index in the redirection vector.  The {\tt vmid} flag identifies
the Container ID associated with the app.  A value of 0 for {\tt vmid}
identifies the trusted host. Higher values identify the application's
container.

This {\tt vmid} value is added as a one byte field in the process descriptor in
the host kernel so that a process's container can be identified by the
system.  This does limit Anception to at most 256 containers, 
but that is likely to be more than sufficient in practice.

When an app runs, the Anception kernel module obtains the
\texttt{task\_struct} of the process, and sets the virtualization byte
({\tt vmid}) to the appropriate index value. The vmid is inherited by
a process's children. Forking/cloning does not allow a process to
escape its container. Also, since this process descriptor is
maintained on the host side, compromise of a root service in the
container does not allow the app to change its {\tt vmid} and escape
the container.

\subsection{Container Implementation}
Containers run a special version of Android that contains the stock
Android system in a guest virtual machine.  It has all the core
userspace system services found in a regular Android system. The
window manager and input manager are nullified by making them talk to
dummy devices. Thus, the capability to render and process touch input
is nullified. We refer to this as {\em headless} Android. 
The point of running this in the guest VM is to provide an alternate,
mostly-identical system view to an untrusted process.

Note that apps are not installed inside container images. Thus,
all container images are essentially identical, except for state
of the apps that are bound to the containers. Furthermore, most
of the runtime state of apps resides outside the container, since
the memory pages of apps mostly reside in the host (except for 
much smaller state of the corresponding proxy that runs in the container).

\subsection{Proxies in an Anception Container}
An app's system call must be forwarded to its proxy (which runs in a guest
container). When a proxy is first started, it is 
granted the same UID (among other parameters like umask). When a redirected system call arrives at the proxy, we need
an efficient way of delegating its execution to the proxy. This means we have to transfer execution from kernel context to the proxy's process
context. One could have a message passing mechanism that notifies a userspace process upon receipt of a system call from the host, but this
would require four costly context switches to complete. 

Anception uses a more efficient technique that eliminates two of the
context switches. We force the proxy process to enter an
\textit{interruptible sleeping wait} in kernel space through an ioctl
on its container's kernel module. When a redirected system call comes
in, we post this call to the container's kernel, which in turn wakes up the 
target proxy. After execution of the system call is complete (without leaving 
the container kernel), the proxy goes back to sleep, in kernel space. 
This mechanism saves us two context switches over the
other option of having the proxy wait in userspace for a notification
from the container's kernel that a system call is to be executed.

\subsection{Efficient System Call Interception}

To be able to redirect system calls, one must first intercept a system
call. We tried out several system call interception mechanisms --  in particular
{\tt ptrace}, {\tt kprobes}, {\tt ftrace}, {\tt dtrace}, and {\tt utrace} --
 but found them to be too slow or limiting. For example, Anception's first prototype was built using \texttt{ptrace} and User-Mode Linux \cite{uml}.
In this experiment, we bound untrusted processes by ptrace-ing them. However, microbenchmark tests indicated overheads upwards of
60x (primarily due to a high number of context switches). 
In Android, all applications derive from Zygote. During a ptrace session, reparenting occurs, and this may cause unintended side-effects. Other mechanisms,
primarily debugging or tracing mechanisms, were also not efficient
enough for our purposes.

We decided to design our own system call interception mechanism that was 
suited for transferring the call to a proxy process in a different container
efficiently.
The system call interception mechanism incurs an overhead of just a few
instructions on the host system.

The basic insight was that we could create a different system call
table per container and then provide an efficient way of redirecting a
system call to the appropriate table, based on the app's container ID
(which is available in the {\tt vmid} value in the process
descriptor as described in part B).

\begin{figure}[ht!]
\center
\includegraphics[width=3.3in,height=3.6in,keepaspectratio]{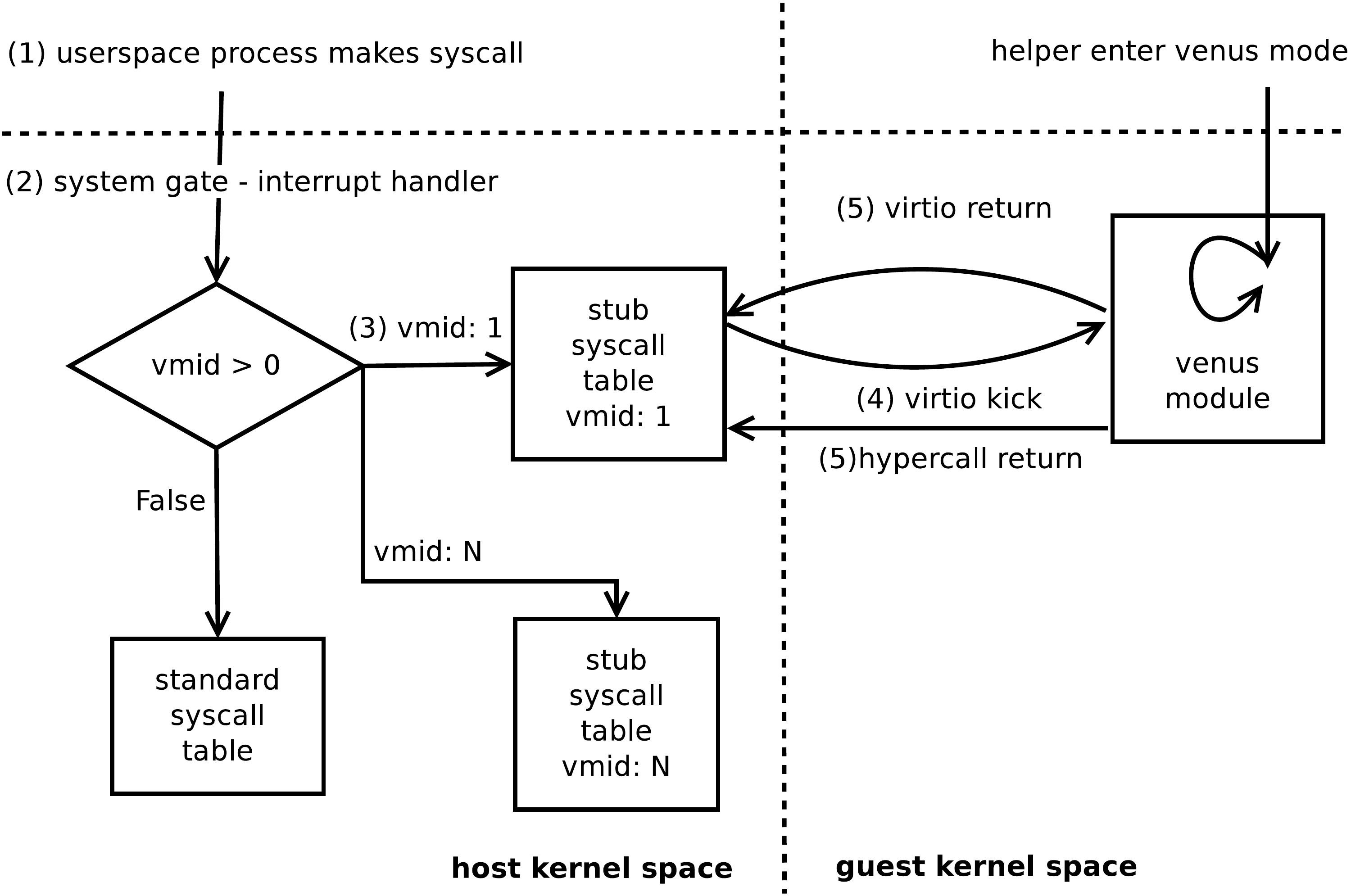}
\caption{Anception Architectural Detail}
\label{arch1detail}
\end{figure}

Figure \ref{arch1detail} shows the details of a system call.  When an
application process makes a system call (step 1), control is
transferred to the system call interrupt handler (step 2). At that
stage, we analyze a field in the process descriptor, named
\texttt{vmid}, that allows us to identify the container that is bound
to the process. A value of 0 indicates that the process runs directly
on the trusted host and no redirection should occur. If {\tt vmid} is
greater than 0, its value specifies an index into a redirection
array. The redirection array is a set of base addresses for various
system call tables corresponding to the various containers. The
implementations of system calls will serialize call arguments (step 3)
and transfer them to the guest kernel via the \texttt{virtio}
transport (step 4).  Once the call has been executed by the host
process's proxy, the results are returned either via \texttt{virtio}
or via a hypercall (step 5). Finally, the results are returned to the
original process that initiated the call.

System call interposition has been the topic of much research in the
past, several problems have been identified
\cite{Garfinkel03trapsand}, and solutions have been developed
\cite{Garfinkel03ostia:a, Venkatakrishnan03isolatedprogram}. Our
system call interposition method follows several best practices
outlined in previous research \cite{Garfinkel03ostia:a,
  Venkatakrishnan03isolatedprogram, Laadan:2010:OSV:1815695.1815717},
while avoiding the problems encountered while building systems such as
in \cite{janus,
  Jain99user-levelinfrastructure}.  Some representative problems
include time-of-check-time-of-use (TOCTOU) attacks, and symlink races.
The choice of which syscalls to intercept has been influenced in part
by the REMUS system~\cite{Bernaschi:2002:RSO:504909.504911}.

Our streamlined implementation of system call interception adds a few
assembly lines to the system call interrupt handler. In practice, this
is efficient. Furthermore, the amount of architectural state we save
to intercept and redirect is negligible. For details, see Table
\ref{archsave}. The minimal runtime overhead of the system call interception is reported
in Section \ref{perf}.

\begin{table}[ht!]
\center
\begin{tabular}{| l | l | l |}

   	\hline
  	entry method		&			save																			&	restore \\ \hline	
  	int 80				  &  movl \%eax, \%ebx															& movl \%ebx, \%eax \\ \hline
  	
  	\multirow{4}{*}{sysenter} & pushl \%esi				 					& popl \%esi \\ 
  														& pushl \%ebx									& popl \%ebx \\
  														& pushl \%ecx									& popl \%ecx \\
 															& movl \%eax, \%ebx 					& movl \%ebx, \%eax \\ \hline

\end{tabular}
\caption{Architectural State Saves}
\label{archsave}
\end{table}

\vspace{-0.2in}

%% file: functional.tex
\section{Security Evaluation}
\label{sec:evaluation}

We now analyze Anception's ability to handle previously existing attacks. All
of the attacks discussed below would be confined in a fully virtualized
system. Anception potentially presents a slightly larger attack surface
for compromising the trusted host than with full virtualization, 
since in Anception there are some system calls that execute entirely in the host environment. While the compromise of the host is outside of our threat model, it is important to understand
the limitations of that assumption.

\input{vuln}

\subsection{Example Exploit Walkthrough -- Gingerbreak}
\label{exploitwalk}
Anception prevents most rootkit attacks. 
To understand how, we present a concrete example of the Gingerbreak exploit.
Gingerbreak is a local privilege escalation based on a negative
integer array access in \texttt{vold}, the volume manager on Android.
It has been used by a number of malware applications
\cite{gingermaster} as a method to gain superuser privileges and
nullify the Android security model. Below, we summarize the steps
that a malicious app with Gingerbreak takes and the actions that occur when that app is running in an Anception container.
\begin{enumerate}
\item Gingerbreak starts out by 
making a copy of itself by reading \texttt{/proc/self/exe} and writing
to the malicious app's private directory. With Anception, 
the write will be redirected to the app's private directory, which is an 
identically named and configured directory in the app's container. Thus, a copy of 
the exploit's executable will be made in the app's container.
\item   The exploit then proceeds to
its information gathering stage. The first step here is to find the
\texttt{vold} daemon by its process identifier. It does this by
opening \texttt{/proc/net/netlink}. With Anception, this \texttt{open} system call will be redirected 
to the app's container and the exploit will read the container's runtime information of the netlink 
environment. We have an identical environment in the container.
\item The exploit then searches \texttt{procfs} for \texttt{/system/bin/vold}
and makes a note of the corresponding PID. As stated in Section \ref{sec:design}, with Anception, 
the system calls to find the pid will be forwarded to the app's container. 

Note that now, the
Gingerbreak exploit executing on the host has obtained the PID of the
\texttt{vold} executing inside the container.

\item The exploit then proceeds to find the address of \texttt{system} and
\texttt{strcmp} inside \texttt{/system/bin/libc.so}. As applications execute
on the host, and any useful application will use {\em libc}, Anception simply allows
opens and reads to execute on the host for such system code libraries. 

\item The exploit, in the next stage of information gathering,
  attempts to find the Global Offset Table (GOT) start address of
  \texttt{vold}. The Gingerbreak exploit does this by opening the \texttt{vold}
  executable and using the ELF-32 API to parse it. The exploit then 
 looks for the last
  piece of information to find the storage device that \texttt{vold}
  manages. System files are involved in the reading process and as per
  our rules, we let them go through on the host itself since
  these files are read-only.
\end{enumerate}

Coming to the actual privilege escalation, Gingerbreak needs to find
the negative index value to send to \texttt{vold} so as to
achieve code execution.  It uses a brute force approach by trying
values in a range and then scanning the logcat crash logs for failed
attempts. The exploit creates its own logcat log file (which is
redirected and created in the app's container), kills logcat (which is mirrored in the app's container as well), and then restarts it by specifying its own 
file as the log file (also restarted in the app's container).  Note that as per
Anception's rules, when a fork/exec occurs, we simply let the fork
happen on the host, but the new process is bound to the app's container; 
the sandbox is extended to the forked process.  
Since the new logcat is bound to app's container, it
sends its output to a file that only exists in that container.

Once an index has been calculated, Gingerbreak forms a netlink message
and uses socket calls to talk with the \texttt{vold} process.
With Anception, Gingerbreak sends shellcode with the negative index to
\texttt{vold} inside the container. This causes \texttt{vold} to execute
the exploit binary that was copied into the container. The exploit always
checks on execution whether its uid is 0. Since \texttt{vold} started
it the second time in the container, the root check succeeds and the
exploit has \textit{succeeded inside the container VM}. As the container is 
isolated via system virtualization, the host remains protected.

%% file: vuln.tex
\subsection{Vulnerability Study}
\label{subsec:vuln}
We examined the past reported vulnerabilities on Android in the Common
Vulnerabilities and Exposures (CVE) system to determine how Anception
compares with running untrusted apps in full virtualization at
protecting against the vulnerabilities.  The CVE system returned 362 CVE
entries on searching for the string ``Android'' from the past four
years. Of these, we found 62 CVEs that had sufficiently detailed
information for analysis, involved exploits of apps, and were deemed
to be protected by full virtualization. There were additional CVEs
(mostly involving Adobe Flash) that lacked sufficient detail on the
precise attack vector. It is likely that both full virtualization and
Anception would protect against those if the vulnerable code was
running as part of an untrusted app, but lack of detail on the exploit
prevented a scientific analysis.

\begin{table}[ht!]
\begin{tabular}{| l | c | c | c |}
\hline
\multirow{2}{*}{Vulnerability Type} & Prevented or    & Prevented or 		   & Prevented or    \\
				    & Confined 	      & Confined by Full	   & Confined 	     \\
				    & by Anception    & Virtualization		   & by Cells	     \\
\hline                                                                                               
Application			    & 47	      & 47	    		   & 47	      	     \\ \hline
System Services			    & 7 	      & 7 	     		   & 7 	      	     \\ \hline
Device Drivers			    & 7		      & 7			   & 7		     \\ \hline
Kernel Core			    & 1 	      & 1 	     		   & 1 	      	     \\ \hline
\hline                                                                                               
Total				    & 62	      & 62	     		   & 62	      	     \\ \hline
\end{tabular}
\caption{Results of manual analysis of CVE records.}
\label{tab:vuln_study}
\end{table}


Table \ref{tab:vuln_study} shows the details of the vulnerability study on the
62 vulnerabilities that we deemed to be preventable when running a malicious
app under full virtualization. Of these, 47 involved apps that did not
adequately protect files or interfaces with other apps or the network, allowing
an attacker to leak sensitive information by gaining full or partial control of
the app.  We judged those vulnerabilities to be preventable in Anception if
the apps were appropriately assigned to the Anception containers and
appropriate policies for network communication governed those containers.

Seven involved system services that could be manipulated by malicious
apps to escalate to root privileges or obtain sensitive system
information (see Section \ref{exploitwalk}).

Another seven were serious Linux device driver vulnerabilities.  These
included four instances of poor bounds checking on memory operations
within graphics and diagnostics drivers. A further two vulnerabilities
were instances of improper dereference and use of pointers to user
space that were passed as arguments to \texttt{ioctl}.  In these six
cases, the vulnerable driver code is accessed by first opening the
device file corresponding to the vulnerable driver and then performing
an \texttt{ioctl} system call to send a request to the driver, which
would contain the exploit payload.  In our experience, there would
never be any reason for an ordinary app running on Android to access
any device files directly other than {\tt binder} and {\tt ashmem}, both of
which are vital services intended to be used by all applications, e.g.,
for sending intents and shared memory. Anception redirects any \texttt{open} 
system calls for drivers other than these two to the guest container, in 
turn confining the vulnerability to the guest container. 

On Cells, there is a higher risk that the above device driver
vulnerabilities will not be confined. There is only one kernel in
Cells shared by all the containers. If the device driver in question
is virtualized as well or disabled in a cell, then there may be a 
possibility that the exploit could fail. 

Finally, there was one instance of a driver file that provided read and write
access to physical memory and was configured as world readable and writable by
the vendor. Again, Anception would prevent it since read/writes to such files
would be redirected to the container.

Finally, one vulnerability in the Linux kernel allowed an attacker to compromise
physical memory by triggering a null dereference in code related to network
socket operations.  Since Anception redirects these system calls, they would
only compromise the container that a malicious app was bound to.

Anception does present a different attack surface than full virtualization
and Cells. Since the display-related activity takes place directly on the
host, bugs in that could potentially be used to exploit Android. We were
unable to find such exploits.

%% file: performance.tex
\section{Performance Evaluation}
\label{sec:experiments}

Anception's design decision is to launch the applications from the host rather than within
the container and not direct graphics operations. To verify that it is the right design
choice for Android,  we conducted a small experiment to measure the relative frequency of
UI updates and touch events as compared to other class of operations. 
We ran five regular applications
available on almost every phone - Email, Web Browser, Music, MMS,
Default Game.  As we see in Table \ref{tab:ioctlstats}, over 92\% of
the system calls made are UI-related.  A similar study
measuring categories of system calls was conducted by ProfileDroid
\cite{Wei:2012:PMP:2348543.2348563} as well. The results from both
studies serve to confirm our hypothesis.

\begin{table}[ht!] \center
\begin{tabular}{| l | c | c |}
\hline
Service Requested			& Total Count	& Percent	\\ \hline
\hline
android.accounts			& 2		& 0.00\%	\\ \hline
android.app				& 1769		& 2.96\%	\\ \hline
android.content				& 1611		& 2.69\%	\\ \hline
android.media				& 16		& 0.03\%	\\ \hline
android.net				& 4		& 0.01\%	\\ \hline
android.os				& 122		& 0.20\%	\\ \hline
android.ui				& 48646		& 81.35\%	\\ \hline
android.utils				& 919		& 1.54\%	\\ \hline
android.view				& 2004		& 3.35\%	\\ \hline
com.android.internal.telephony		& 82		& 0.14\%	\\ \hline
com.android.internal.view		& 4619		& 7.72\%	\\ \hline
ImountService				& 1		& 0.00\%	\\ \hline
\end{tabular}
\caption{Counts of various Binder IPC ioctls issued to Android services.
Note that UI dominates other types of requests.}
\label{tab:ioctlstats}
\end{table}

Next, we run a microbenchmark to analyze the overhead of the VM-independent system call interception framework.
We run two popular Android macrobenchmarks to analyze whole system performance and one application level
benchmark to analyze the impact on end users. The macrobenchmarks are used so that we can measure the performance
for interactive workloads that are the most common for smartphones. We also report on the memory consumption
of the guest VMs.

All the benchmarks are run on an Asus EEEPC ET1602 equipped with an Intel Atom x86 processor,
1GB of RAM and 160GB of internal flash memory. The EEEPC runs Android x86 version 2.3.7 with Linux kernel version 2.6.39.
We configured the guest VM with 64MB of main memory that ran our headless Android port.

\subsection{Microbenchmarks}
\label{perf}
We run a microbenchmark to measure the latency of performing 3 common system calls with the Anception Syscall Interception Method (ASIM) activated.
We measure the time for a \textit{null-redirection}, which is intercepting a syscall and
redirecting it to the real system call handler.
The benchmark executes the syscalls 10,000 times and uses the x86 cycle counter for timing.
Trial runs are executed by the benchmark to warm up the caches. The results for the ASIM are summarized
in Table \ref{asim}. The write, read calls are executed on files.

\begin{table}[ht!]
\center
\begin{tabular}{| l | c | c |}

   	\hline
   	syscall							&							Native ($\mu s$)						& 					Anception ($\mu s$) \\ \hline
   	getpid							&						0.57									& 					0.57 \\ \hline
   	write (32 bytes)		&						17.7								&						17.8 \\ \hline
   	read (32 bytes)			&						3.3									&						3.4 \\ \hline
\end{tabular}
\caption{ASIM latency}
\label{asim}
\end{table}


As we can see in the table, the ASIM adds negligible overhead and is very efficient.
The next microbenchmark measures the round-trip time for a system call when redirected to an active guest VM.
The results are summarized in Table \ref{rtt}.

\begin{table}[ht!]
\center
\begin{tabular}{| l | c | c |}

   	\hline
   	syscall							&							Native ($\mu s$)						& 					Anception ($\mu s$) \\ \hline
   	write (16 bytes)		&						24.9								&						206.9 \\ \hline
   	read (16 bytes)			&						3.4								&						203.2 \\ \hline
   	write (32 bytes)		&						17.7									&						196.6 \\ \hline
   	read (32 bytes)			&						3.3									&						212.7 \\ \hline
   	
\end{tabular}
\caption{Full redirection round trip time}
\label{rtt}
\end{table}


The increase in latency is caused due to two VM switches (host $\rightarrow$ guest, guest $\rightarrow$ host) and process context switches.
As context switch times go down due to improvements in hypervisor technology, we expect to see reductions in the latency. However, these numbers do not adversely affect performance
at a macro level, which is what we report on next.

\subsection{Macrobenchmarks}
We run three popular Android macrobenchmarks -- PassMark \cite{passmark}, AnTuTu \cite{antutu} and SunSpider \cite{sunspider}. These benchmarks are designed to test 2D and 3D graphics
performance, Disk I/O, Memory I/O and CPU (Java, Native) performance. They report a weighted total score, and individual scores.
Running 2D/3D benchmarks enables us to analyze the overheads caused by our \texttt{ioctl} handling logic. It should not cause
a large overhead because graphics related \texttt{ioctls} form a large chunk of the system 
calls executed by applications (Table \ref{tab:ioctlstats}).

\begin{figure}[ht!]
\center
\includegraphics[width=3.6in,height=3.5in,keepaspectratio]{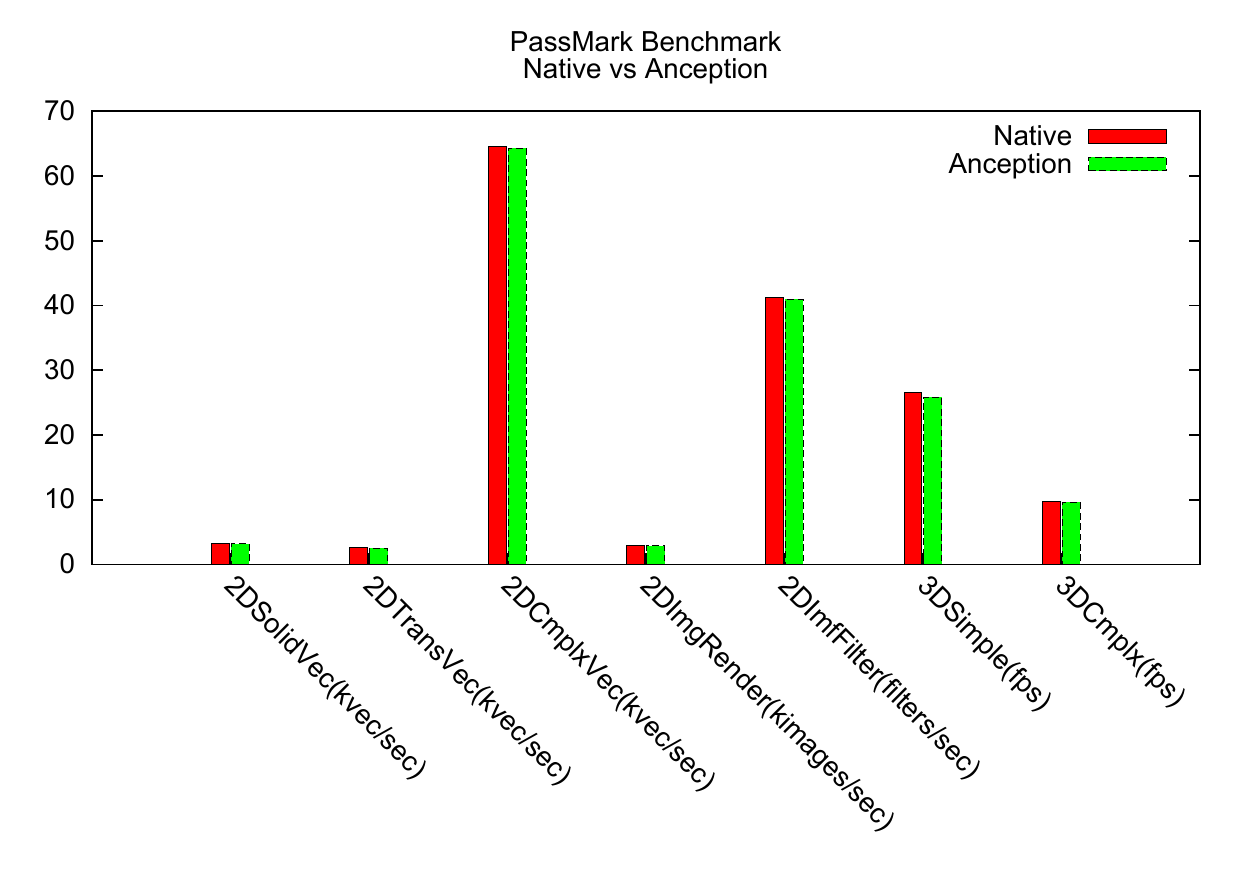}
\caption{PassMark Benchmark}
\label{passmark}
\end{figure}

The PassMark benchmark reports details of the tests conducted. Figure \ref{passmark}
shows us that Anception incurs negligible overhead for a range of 2D and 3D graphics tests.
They also show no noticeable change. Overhead of 3.88\% is incurred on 2D tests and 2.64\% on 3D tests. In terms of CPU and Memory I/O, as reported by PassMark,
Anception incurs 0.43\% overhead on memory operations and 2.3\% overhead on CPU operations.
This slight increase is due to the guest VM running Headless Android that puts additional scheduling and 
memory pressure on the host. However, due to our architecture of handling UI operations on the host,
the impact of this VM is negligible. 

\begin{figure}[ht!]
\center
\includegraphics[width=3.6in,height=3.5in,keepaspectratio]{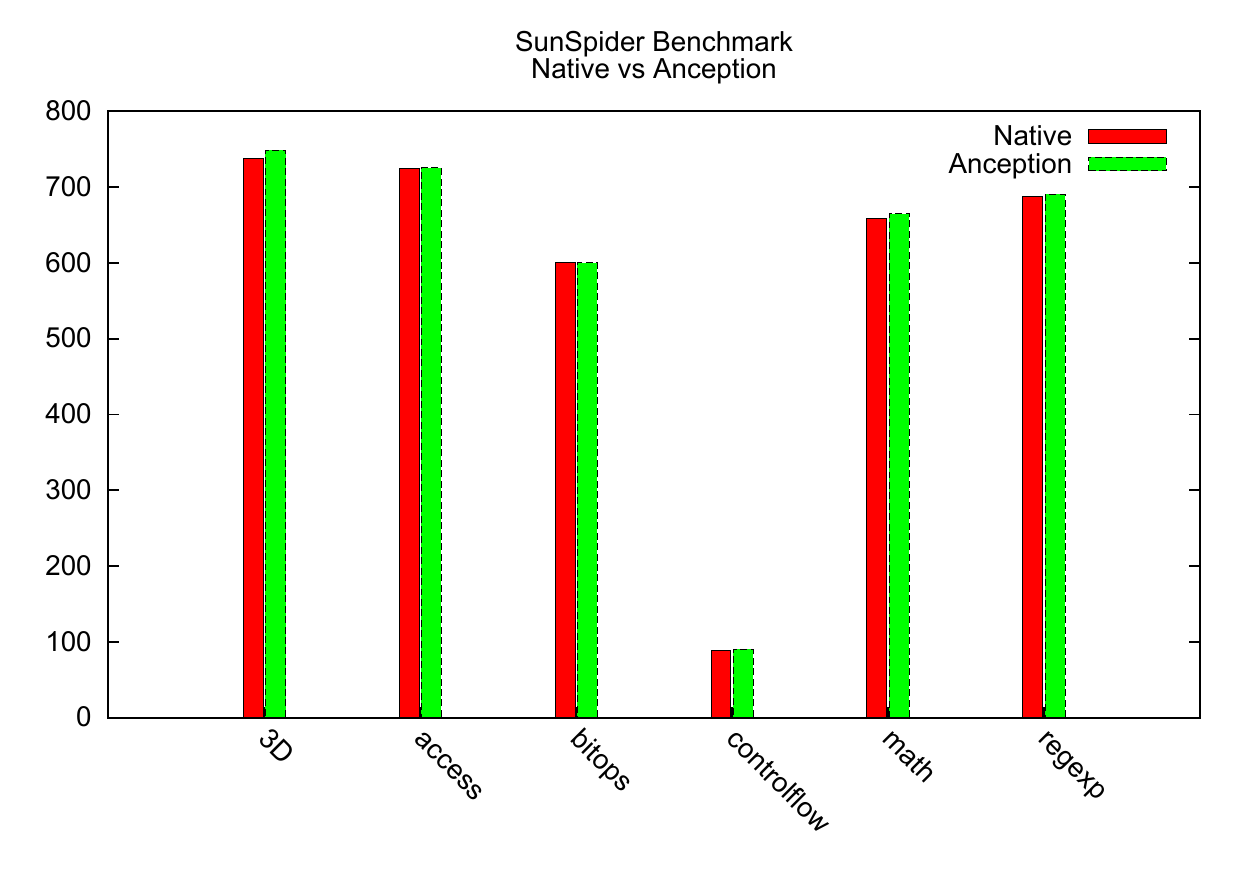}
\caption{SunSpider Benchmark -- Time in milliseconds}
\label{sunspider}
\end{figure}

On the AnTuTu Benchmark, Anception's score was 3.6\% less than native, with a 0.4\% overhead on the 2D tests and 1.3\% overhead on the 3D tests.
To get an idea of Anception's performance with interactive and application oriented workloads,
we run the SunSpider \cite{sunspider} Benchmark. At a macro level, from Figure \ref{sunspider}, we observe
that Anception closely follows the performance of native execution with a 1.2\% maximum overhead among all tests. This benchmark also shows our ability to run unmodified
Android applications on Anception. The reason for the modest overheads is the \texttt{ioctl} identification code wherein we
determine whether a particular IPC operation is related to graphics and input operations.

\subsection{Memory overhead}
Finally, we quantify the memory (RAM) savings earned by using a headless version of Android. To measure this,
we read the \texttt{/proc/meminfo} file on a stock Android 2.3.7 system, and we read the meminfo file
on a headless instance. These readings were taken when the systems were just booted up and just the bare essentials
were running.
For a stock Android system, the number of active bytes of memory is 99.11 MB.
On the headless version, it is 14.87 MB, or 15\% of stock Android. This is much smaller compared
to 40\% of stock Android that is needed by Cells to start a guest VM. Moreover, we have not applied
optimizations like Kernel Samepage Merging (KSM) that Cells has applied to its implementation. We expect
to see further reductions with KSM in place.
The reason for the decrease is again due to Anception's architecture.
By running a Headless Android instance in the guest VM, the memory used up by Display Managers and Input Managers
is freed up, allowing us to run smaller sized (in terms of RAM) VMs. However, during all our experiments, we started
the guest VMs with 64MB of main memory, although we could have started them with as low as 44MB of memory. We did not
do this as we didn't want to adjust the Android low memory killer parameters.

\subsection{Power Measurements}
Smartphones are resource constrained devices and usually have limited battery power. In order to ensure that the
effect on energy consumption due to Anception is not significant, we conducted several experiments on the PassMark and SunSpider benchmarks to quantify 
the overhead. We used a Watts Up? .Net power meter for our measurements on power and energy consumption. Using the \textit{MonkeyRunner} 
feature provided by the Android platform, we built automated tests to run the benchmarks for an hour each, both with 
and without Anception enabled. In order to simulate the actual execution of an application, the benchmark was exited using the back button 
after every run. For the PassMark benchmark, we observed that the energy consumption was 20.8 Watt hours without binding it to Anception 
and the consumption went up to 21.3 Watt hours upon binding, giving an overhead of 2.4\%. For the SunSpider benchmark, energy consumption 
without Anception was 34 Watt hours and on binding the consumption went up to 35.3 Watt hours indicating an increase of 3.82\%.

%% file: limitations.tex
\section{Limitations}
\label{sec:limitations}
As Anception is a research prototype, we acknowledge a few limitations of the system. 

As we described in Section~\ref{sec:design}, intent requests to all
Android system services, except for the touch display, are directed to
the container. If an app sends a request to the Location Manager to
get the GPS readings, the IPC request will go to the guest and then be
delivered to the Location Manager in the container.  That Location
Manager will talk to a virtual GPS device. Then, it is a matter of
policy whether the virtual device should be permitted to talk to the
real device on the host via a narrow interface.  Our research
prototype has not implemented these mappings from virtual devices
to physical devices, but the technology to do that is well understood.
Most virtual machines, such as VMware, have that capability.

The Linux kernel allows certain system calls to be implemented by
drivers. Examples include the \texttt{ioctl} or VFS system calls
(\texttt{read, write}).  Since these are relatively new
implementations, they can be buggy.  As Anception does not redirect
all system calls all the time, it can happen that some system calls
that go to such buggy drivers are executed on the host. In such cases,
Anception is unable to provide sufficient isolation so as to protect
the integrity of the host. However, this is an area of current
research and we could build upon other solutions in the literature
designed to tolerate buggy drivers \cite{virtuos}.

UI system calls and related non-redirected system calls could
themselves contain bugs. As such calls are always executed on the
host, there is a small chance that they could be exploited by
malicious apps. However, given that such calls are heavily used, it is
likely that most errors have been corrected.

Anception only supports memory-mapped file I/O partially. If a file is in the read-only part of the
file system (e.g., /system on Android), those can be memory-mapped since the I/O operations on those
are directed to the host. Memory-mapping in the writeable portion of the filesystem is not supported.
Those files reside in the container. This is not a fundamental architecture limitation. NFS-mounted
files can be memory-mapped. A possible strategy for implementing that would be to provide an NFS-style
access for the host to a container's filesystem.

Anception does not currently use kernel same page merging
optimizations~\cite{ksm}.  The results from the Cells~\cite{cells}
paper show that those optimizations can significantly reduce the
memory requirements for multiple containers. We expect containers to
have a high degree of overlap in pages since, except for the small
proxies to hold system-relevant state, the memory pages of an
application reside on the host.

Anception currently only supports the simple policy that only apps
having the same UID are assigned the same container. It would be
interesting to explore other types of policies. For example, all
work-related apps could be assigned to belong to one container and
personal apps belong to another container. In that case, Anception can
be used as a platform to support multiple virtual phones, as in
Cells~\cite{cells}.

%% file: related.tex
\section{Related Work}
\label{related}

We categorize the related research into several classes and draw on ideas and techniques from several of them.

\subsection{OS Virtualization}
Systems based on paravirtualization such as VMware~\cite{Barr:2010:VMV:1899928.1899945}
Xen~\cite{xenarm} as well as Cells~\cite{cells} can provide a a full virtual
container for the entire operating system.  Cells, in particular, is a
good solution for virtualization on Android if the goal is to support
multiple virtual phones. It uses a single shared kernel. The goal of
Anception is to provide application-level encapsulation. It may be
possible to provide Cells-based Anception where, instead of using
lguest, cells are used to provide containers. This would require a
substantial redesign as, currently, each cell provides a full Android
environment with its own set of apps. Conversely, it may be possible
to provide Cells-like virtual phones using Anception as a platform. It
will require significant changes to Anception's handling of IPCs
(cross-container IPCs will need to be forbidden) and apps will need to
be allowed to be associated with multiple containers.

The recent ExpressOS system \cite{Mai:2013:VSI:2451116.2451148} by Mai
et al is an OS built in a type-safe language. The aim is to build a
small verifiable OS kernel that is able to run apps with high
assurance of the services provided to them. ExpressOS provides a secure
storage service (among others) that guarantees data written to disk is
not tampered with or disclosed to an untrusted Android runtime. The
difference with Anception is that ExpressOS does not strengthen the
isolation between trusted and untrusted apps. If an application is
benign but buggy, ExpressOS does not provide protection against
exploitation by other apps. ExpressOS is complementary to Anception in
addressing a different threat model.

VirtuOS \cite{virtuos} by Nikolaev and Back is a modified Linux environment
that isolates device drivers in virtual machine containers, which run in
parallel to improve performance on multicore environments. Anception does not
address device driver security. VirtuOS is complementary to Anception in
addressing a different threat model.

Anception builds on techniques from several systems.  The Pods system
\cite{Laadan:2010:OSV:1815695.1815717} uses system call interposition as the
mechanism for virtualizing a process's view of the OS.
Linux Containers is another mechanism being integrated into the Linux
kernel, which will allow a user to establish isolated containers via a
tool called LXC.  Linux Containers use a namespace mechanism to
isolate and virtualize resources directly from within the kernel.  
The interposition mechanisms of Pods and Linux Containers has similarities
to that used in Anception. Both Pods and Linux Containers are full
solutions that handle virtualization of an entire guest system.  In
this way, they are interchangeable with lguest in our implementation
of Anception, but do not include any policy for deciding which system
calls are worth keeping on the host for performance reasons.  Because
of this, we view them as complementary tools to our work.

Overshadow~\cite{overshadow} provides mechanisms for applications to protect
their state from compromised kernels.

\subsection{App Sandboxing}

Providing strong isolation on Android is mostly centered around policy based
approaches. TrustDroid \cite{Bugiel:2011:PLD:2046614.2046624} monitors IPC
channels, network and filesystem accesses through extensive modifications to
the Android framework and prevents apps in different trust domains from talking
to each other. Unlike Anception, an untrusted app in TrustDroid executes
entirely on the host kernel, as there is only one kernel. That potentially
makes TrustDroid vulnerable to native code exploits via system calls that would
be forwarded to the guest kernel in Anception but  are executed on the
host kernel in TrustDroid.

AppFence \cite{Hornyack:2011:TAD:2046707.2046780} prevents access to a user's
private data (e.g., contacts data) by providing false or pre-recorded data to
untrusted apps, based on policy.  Aurasium \cite{Xu:2012:APP:2362793.2362820}
inserts inline reference monitors in untrusted applications and monitors their
interactions through the system calls made. In some cases, it rewrites system
call arguments and return values (at the libc level) to enforce certain types
of policies. 

Janus \cite{janus} is a user-level tool that allows users to sandbox
applications and implement a policy that determines which system calls they are
allowed to execute.  If a particular system call invocation is not expressly
permitted by the user-specified policy, it is discarded and returns an error to
the calling process.  Anception is much more flexible than this, in that it
provides an isolated environment in which risky calls can execute safely,
rather than simply denying them.

Wine Windows Emulator on Linux and Drawbridge for
Windows~\cite{libraryos} provide a linkable user-space OS library that
encapsulates an entire Windows OS to which executables can link to.
Part of the goal of these is to allow applications to be portable
across different versions of Windows (they can be distributed bundled
with the OS library). But, another potential use of Drawbridge, in
particular, is to sandbox individual applications. There are some
similarities in that Anception also allows individual applications to
be sandboxed but there are also some differences in both design and
philosophy. Drawbridge uses Windows Remote Desktop to allow
encapsulated apps to share a common display. Anception's apps have
direct access to the Android display, which, by design, should have
lower latency and be more efficient on mobile devices, especially for
gaming apps that are highly interactive. Anception allows multiple
applications to be assigned to a common sandbox, i.e., Anception
containers. Drawbridge applications cannot
share system state.

\subsection{File System Isolation}

Android recently incorporated a multiuser feature that helps in setting up
multiple user accounts and sharing of a single device. Each user is assigned a
unique user ID and corresponding directory on the filesystem
(\texttt{/data/users/ID}). When the device switches to a user, symbolic links
are set up from an app's directory (\texttt{/data/data/APP.PKG}) to the private
user directory. This enables backwards compatibility and makes a profile switch
efficient, but the overall design has drawbacks. If an owner wants to share an
app, he must first create a secondary profile, switch profiles and reinstall
the app into the new profile. If the owner uses his own Google account, various
settings can be imported into the other profile, causing privacy leaks. Direct
sharing from the primary user profile is not possible, although there are
community built apps that enable such sharing. These have the disadvantage of
sharing all private data from the owner's profile with the secondary profile.

\subsection{Information Flow Techniques}

Moses \cite{Russello:2012:MSO:2295136.2295140} is a system that provides
isolation among apps in different trust domains. The main disadvantage of Moses
is that it requires heavyweight taint tracking at run-time to detect and
prevent undesirable information flows.  Moses uses filesystem virtualization
(namespaces) to provide a different data directory to an app.  This allows an
app to maintain different profiles by switching filesystem namespaces. This
supports sharing of apps easily but does not provide a secure and isolated
execution environment for untrusted apps.

%% file: conclusions.tex
\section{Conclusion}
\label{sec:conclusion}

Current virtualization solutions, although useful, fall short of user
expectations. Anception proposes a headless virtualization model that achieves a
balance between isolation strength, user experience, and system performance
while protecting a user's internal (on-device) and external (online) data.
Through a combination of system call redirection and system virtualization,
Anception provides the technology to run a subset of an app in an isolated,
virtual container while maintaining correctness of execution.  An exhaustive
evaluation of Anception in terms of security guarantees and system performance
revealed overheads of up to 3.88\% on various 2D and 3D benchmarks and up to
3.82\% additional power usage, as well as comparable isolation guarantees to
existing virtualization mechanisms, showing that Anception is indeed a
practical system.  As future work, we aim to investigate other forms of system
virtualization like namespaces, while keeping kernel changes modest and
maintaining security guarantees.

%% file: ack.tex
\section{Acknowledgements}
This material is based upon work supported by the National Science Foundation under Grant Numbers 0916126 and 1318722.
Any opinions, findings, and conclusions or recommendations expressed in this material are those of the author(s) and do not necessarily reflect the views of the National Science Foundation.